\documentclass[aps,pra,superscriptaddress,twocolumn]{revtex4-1}
\usepackage{dcolumn}
\usepackage{amsmath}
\usepackage{amsfonts}
\usepackage{amssymb}
\usepackage{graphicx}
\usepackage{hyperref}
\usepackage{xcolor}

\newcommand{\brm}[1]{\ensuremath{\mathbf{#1}}}

\begin{document}
\title{Partial coherence and coherence length in StimPDC}
\author{G. H. dos Santos}
\email{fisica.gu@gmail.com}
\affiliation{Departamento de F\'{i}sica, Universidade Federal de Santa Catarina, CEP 88040-900, Florian\'{o}plis, SC, Brazil}
\author{R. C. Souza Pimenta}
\affiliation{Departamento de F\'{i}sica, Universidade Federal de Santa Catarina, CEP 88040-900, Florian\'{o}plis, SC, Brazil}
\author{R. M. Gomes}
\affiliation{Instituto de F\'{i}sica, Universidade Federal de Goi\'{a}s, CEP 74690-900, Goi\^{a}nia, GO, Brazil}
\author{S. P. Walborn}
\email{swalborn@udec.cl}
\affiliation{Departamento de F\'{\i}sica, Universidad de Concepci\'on, 160-C Concepci\'on, Chile}
\affiliation{Millennium Institute for Research in Optics, Universidad de Concepci\'on, 160-C Concepci\'on, Chile}

\author{P. H. Souto Ribeiro}
\email{p.h.s.ribeiro@ufsc.br}
\affiliation{Departamento de F\'{i}sica, Universidade Federal de Santa Catarina, CEP 88040-900, Florian\'{o}plis, SC, Brazil}

 \begin{abstract}
In parametric down conversion, a nonlinear crystal is pumped by a laser and spontaneous emission takes place in signal and idler modes according to the phase matching conditions. A seed laser can stimulate the emission in the signal beam if there is mode overlap between them. This also enhances the emission in the idler beam,  affecting its coherence properties. While the degree of coherence of the idler field as a function of the seed power has already been studied, the transverse coherence length has not yet been properly investigated. The transverse coherence length is a key parameter of optical beams that determines the beam divergence, for example. Here, we present a theoretical and experimental investigation of the transverse coherence length in stimulated down conversion. In addition, we make a connection between stimulated down conversion and partially coherent sources like the Gaussian Schell model beams, and show that in general the idler field cannot be described with this model. 
 \end{abstract}

\maketitle

\section{Introduction}
The coherence of a light field is a fundamental property that affects all degrees of freedom.
In several circumstances it is possible to treat temporal coherence and transverse spatial coherence
as independent properties, and here we focus in the spatial coherence. Most natural light sources
are spatially incoherent in the source, but becomes partially coherent with propagation\cite{born_wolf99}.
Therefore, partial coherence is a relevant issue in most applications of natural light. 
A particularly relevant case of partially coherent light is the so called Gaussian Schell Model beam (GSM)\cite{Schell56,friberg94},
which finds applications in imaging, optical communication, light scattering, nonlinear optics, and others \cite{Ma:17,ismail2017,Zhang:19,ismail2020,Hutter20,Hutter21,Santos2022,Cai2022}.

On the other hand, laser fields are nearly perfect coherent. In this case the light can
be described by deterministic electric field functions, while the partially coherent ones require
other mathematical tools that include stochastic fluctuations, as in the GSM model. We analyze here
a different kind of partially coherent light source: stimulated parametric down conversion (StimPDC).
In this nonlinear process, spontaneously emitted light is mixed with stimulated emission in the so-called idler mode. The degree of
coherence of the light emitted depends on the stimulating power \cite{Ribeiro95} and so it is expected
that the coherence properties will also depend on this parameter. We present here a theoretical and experimental investigation
of the coherence length of the light produced in StimPDC. To our knowledge, this coherence property has not yet been
investigated previously. 

 \section{Theory}
 \label{sec:theory}

 The mutual coherence of the idler field in StimPDC with a coherent seed beam was studied in Ref. \cite{Ribeiro95}, where it was shown that:
 \begin{equation}
 \mu_{i} = \frac{\mu_{sp} I_{sp} + \mu_c I_c}{I_{sp} + I_c},
 \label{mustim}
 \end{equation}
 where $\mu_{i}$ is the normalized mutual coherence between fields at points 1 and 2 in a plane transverse to the propagation direction of the idler beam (the position dependence is not included explicitly in $\mu$ for brevity). For StimPDC the mutual coherence is composed of two terms.  The first, which we denote as $\mu_{sp}$, is the mutual coherence between fields at the same points due to the contribution of the spontaneously emitted light, and $\mu_c$ is the contribution due to the coherent stimulation by the seed beam and is ideally unit. The parameters $I_{sp}$ and $I_c$ describe the intensities of the spontaneous emission and stimulated emission, respectively.
 \par
 The normalized mutual coherence $\mu$ can be calculated using the cross spectral density (CSD) $W(\brm{r}_1,\brm{r}_2)$:
 \begin{equation}
     \mu \equiv \mu(\brm{r}_1,\brm{r}_2) = \frac{W(\brm{r}_1,\brm{r}_2)}{\sqrt{W(\brm{r}_1,\brm{r}_1)W(\brm{r}_2,\brm{r}_2)}}.
     \label{eq:ufromW}
 \end{equation}

\par
As mentioned above, the output idler beam has components originating from the spontaneous and stimulated emission at the source.  It has been shown recently that, in typical experimental situations,  both of these contributions can be described using the well-known GSM beams, which have a cross-spectral density (CSD) given by
\begin{equation}
W_{GSM}(\brm{r},\brm{r}^\prime) = I e^{-\frac{r^2 + r^{\prime 2}}{4 w^2}} e^{-\frac{(\brm{r} - \brm{r}^{\prime})^2}{2 \delta^2}} e^{-ik \frac{(\brm{r}^2 - \brm{r}^{\prime 2})}{2 R}} ,
\label{eq:tgsm}
\end{equation}
where $I$ is the maximum beam intensity, $w$ is the beam waist, $\delta$ the transverse coherence length, and $R$ the phase curvature. 
\par

In the case of the spontaneous emission term and nearly degenerate signal and idler fields, Hutter et. al. \cite{Hutter20,Hutter21} showed that the idler beam produced in SPDC with a GSM pump beam is described by a GSM beam. Thus, we can define the spontaneous (``sp") component with CSD $W_{sp}=W_{GSM}$ for some specific parameters $I, w,\delta,R$.   
The stimulated (``stim") component, on the other hand, has been studied in Ref. \cite{Santos2022}. 
Considering both the seed and pump as GSM beams, the CSD of the stimulated component can be written as $W_{stim}=W_{GSM}$ for some intensity depending also on nonlinear coefficients of the crystal, and the following 
idler beam parameters given as functions of the pump and seed beam parameters:
\begin{equation}
\frac{1}{w_{stim}^2} = \frac{1}{w_{seed}^2} + \frac{1}{w_{pump}^2},
\label{eq:wi}
\end{equation}
transverse coherence length 
\begin{equation}
\frac{1}{\delta_{stim}^2} =  \frac{1}{\delta_{seed}^2} + \frac{1}{\delta_{pump}^2},
\label{eq:deltai}
\end{equation}
phase curvature
\begin{equation}
\frac{k_i}{R_{stim}}= \frac{k_{pump}}{R_{pump}}-\frac{k_{seed}}{R_{seed}},
\label{eq:Ri}
\end{equation}
 where $stim,seed,pump$ stand for stimulated idler, seed, and pump, respectively.

Including the contribution coming from both the spontaneous and stimulated components in StimPDC,
the CSD for the idler field becomes:
\begin{equation}
W_i(\brm{r}_1,\brm{r}_2) =  W_{stim}(\brm{r}_1,\brm{r}_2) + W_{sp}(\brm{r}_1,\brm{r}_2).
\label{eq:Wi}
\end{equation}
 Both $W_{stim}$ and $W_{sp}$ can be GSM-type, depending on the pump and seed CSD and relative intensities. 
 Therefore, $W_i$ will be GSM-type in several practical conditions.
 However, in general it may not be a GSM beam. 
 \par
 Dividing Eq. \eqref{eq:Wi} by the total intensity $I_i=I_{stim}+I_{sp}$ leads to a relation for the normalized mutual coherence of the total idler field: 
 \begin{equation}
 \mu_i = (1-\beta)\mu_{sp}+\beta\mu_{stim},
 \label{eq:mui}
 \end{equation}
 where $\beta = I_{stim}/(I_{sp} + I_{stim})$.
 \par
Let us consider the case where  pump and seed beams are pure Gaussian beams, so that $\delta_{pump} = \delta_{seed} \rightarrow \infty$, giving a stimulated component which is completely coherent $\delta_{stim}\rightarrow \delta_c=\infty$.  In this case, the normalized mutual coherence from Eq. \eqref{eq:mui} leads directly to \eqref{mustim} with $\mu_{stim} = \mu_c$, as observed years ago. 
For the GSM fields, the normalized coherence is a function of the relative distance $d=|\brm{r}_1-\brm{r}_2|$ between points 1 and 2. However, if we assume that the stimulated component is coherent,  $\mu_{stim}=\mu_c$ is constant, ideally unity and independent of $d$. Thus, we can rewrite the mutual coherence of StimPDC as
 \begin{equation}
 \mu_{i}(d) = (1-\beta) \, \mu_{sp}(d) + \beta \, \mu_c,
 \label{mu_d_stim}
 \end{equation}
 where now $\beta = I_c/(I_{sp} + I_c)$. Typically, the transverse coherence length $l_t$ is defined as the value of $d$ for which the absolute value of the normalized mutual coherence is $|\mu_{i}(d = l_t)| = \epsilon$, where $\epsilon$ is a small arbitrary constant. 
 
 The coherence length can be obtained by solving the following relation for $l_t$ and some specific function $\mu_{sp}(x)$:
 
  \begin{equation}
 |\mu_{i}(l_t)| = |(1-\beta) \, \mu_{sp}(l_t) + \beta \, \mu_c| = \epsilon.
 \label{mu_l_t_stim}
 \end{equation}   
 
 Before we further discuss the characterization of the coherence length, let us present an experiment and results investigating the mutual coherence of StimPDC.

\section{Experiment}
\begin{figure}[h]
    \centering
    \includegraphics[width=0.9\columnwidth]{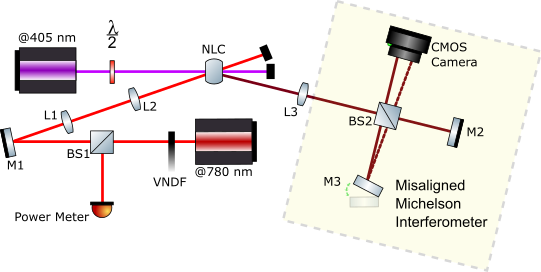}
    \caption{Experimental setup for measuring the coherence length StimPDC light. @405 is a blue laser at 405 nm, @780 is a red laser at 780 nm, $\frac{\lambda}{2}$ is a half-wave plate, VNDF is a variable neutral density filter, NLC is a nonlinear crystal, M1, M2, and M3 are mirrors, BS1 and BS2 are beam splitters, L1, L2, and L3 are lenses. A CMOS Camera measure the interference pattern of the idler beam in a controlled misaligned Michelson interferometer.}
    \label{fig:setup}
\end{figure}

The experimental setup is sketched in Fig. \ref{fig:setup}. A $405$ nm diode laser is used to pump a 2 mm long Beta-Barium-Borate (BBO) nonlinear crystal (NLC), where parametric down conversion takes place. The pump beam is linearly polarized along the horizontal direction and the phase matching is type I. A second diode laser called {\em seed} oscillating at $780$ nm is aligned with one of the down converted modes, say the signal beam, and its emission is stimulated. A scheme using a variable neutral density filter (VNDF), a beam splitter (BS1) and a power meter is used to control the intensity of the seed beam. The lenses L1 and L2 are used to control the size beam on the nonlinear crystal.  The corresponding idler beam at $840$ nm is indirectly stimulated, due to the photon-pair emission in the process. In the idler beam, a lens L3 transposes the image plane to a Michelson interferometer, which is 8 cm away from the crystal. The strategy to measure the transverse coherence length consists in misaligning the interfering beams from the interferometer shifting one path by a tilt in the mirror M3 so that different points in the wavefront are superposed with different separations as illustrated in Fig. \ref{fig:setup}.

\section{Results and Discussion of Coherence Length} 

\begin{figure}[h]
    \centering
    \includegraphics[width=1.0 \columnwidth]{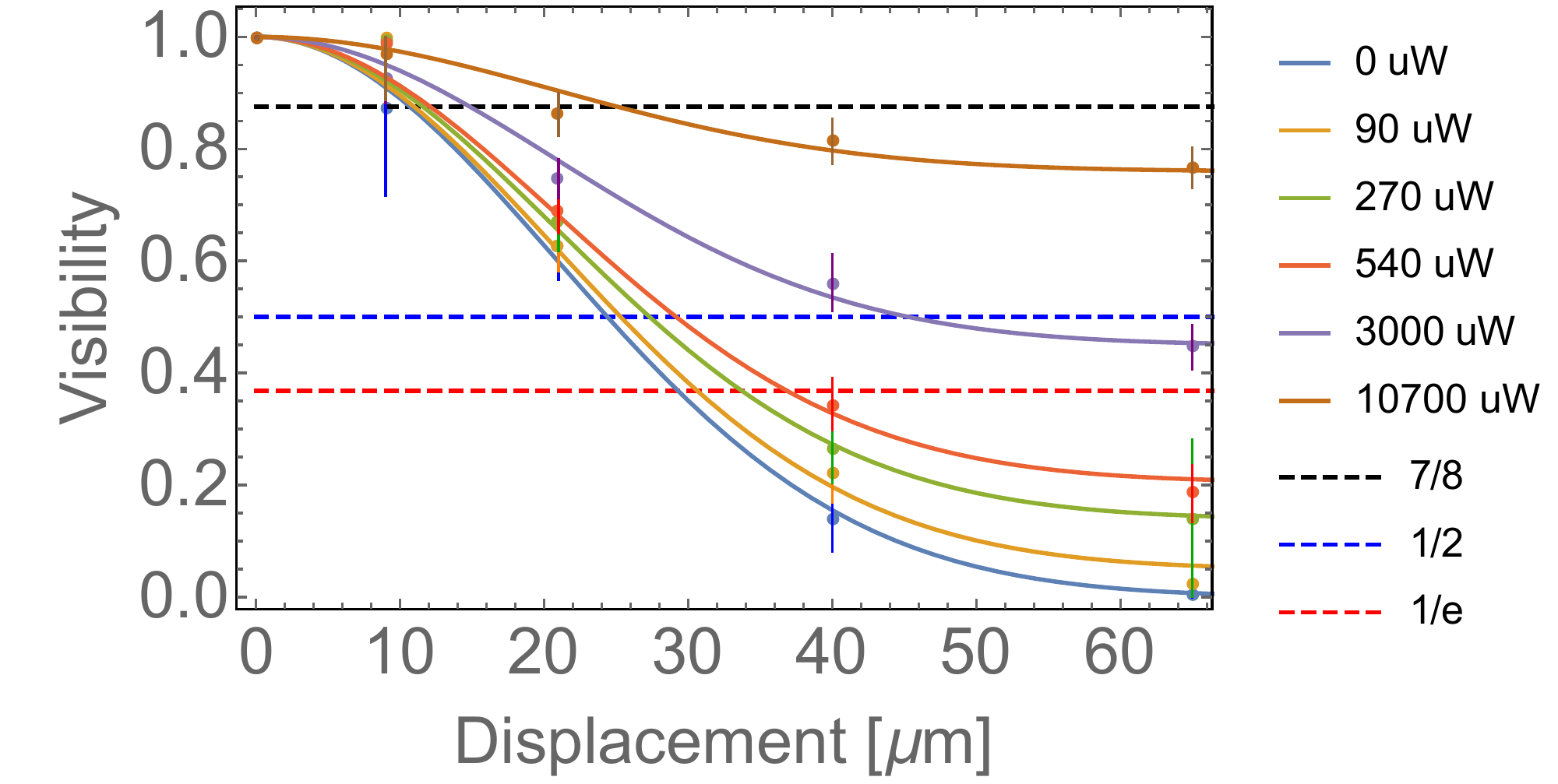}
    \caption{Mutual coherence as characterized by the visibility of the interference pattern as a function of beams displacement $d$ in a misaligned Michelson interferometer}
    \label{fig:res1}
\end{figure}

Fig. \ref{fig:res1} shows plots of the mutual coherence, identified as the visibility of interference patterns for different displacements $d$. They were obtained varying the separation between the interfering beams at the output of the Michelson interferometer. The separations are on the order of tens of microns, and are controlled by tilting one of the beams and measuring the displacement between the peaks of the two interfering distributions taken separately.
The interference patterns are photographed with a CMOS camera. Measurements are performed for several different separations and several different ratios between the spontaneously emitted and the stimulated emitted light contributions.  

Qualitatively, we observe that the visibilities of the interference patterns decrease with increasing separation between interfering beams, as expected. 
Moreover, the degree of coherence for  displacement $d>0$ always increases when the seed power increases. 
\par
Using Eq. \eqref{eq:ufromW} and the GSM for the spontaneous contribution, we can calculate $\mu_{sp}(d)=e^{-\frac{-d^2}{2 \delta_{sp}^2}}$. The normalized mutual coherence of the idler field is then \begin{eqnarray}
 \mu_i(d)=(1-\beta) \, \mbox{e}^{-\frac{(\brm{r} - \brm{r}^{\prime})^2}{2 \delta_{sp}^2}} + \beta.
 \label{l_t_stim}
\end{eqnarray}
 The absolute values in Eq. \ref{mu_l_t_stim} were dropped, because all quantities are positive.
The data points were adjusted to Eq. \eqref{l_t_stim}. With $\beta=0$ we found $\delta_{sp}=20.72  \pm 0.59\mu\mathrm{m}$, and used this value for all the curve fits so that only $\beta$ was the only free parameter, the values of which are shown in Table \ref{tab:1}.
As seen in Fig. \ref{fig:res1} and also in Eq. \eqref{l_t_stim}, one signature of the light emitted in StimPDC process is that the degree of coherence does not always decrease tending to zero.
The minimum value of the degree of coherence is given by $\beta$, the normalized intensity of the stimulated component, since the stimulated light is nearly as coherent as the pump and seed. 
Therefore, if one characterizes coherence length using a minimum value of the mutual coherence as in Eq. \eqref{mu_l_t_stim}, there may be partially coherent beams (those with $\beta > \epsilon)$
with infinite coherence length, in principle.
\par

At the same time, the choice of parameter $\epsilon$ is arbitrary.  To have some operational meaning, it should correspond to the coherence requirements for a particular application.  For example, in quantum information, certain conclusions about the quantum nature of the source can only be reached for a minimum value of the visibility \cite{borges10,carvalho12,borges14}.  Thus, in order to characterize the coherence length of a StimPDC light source, we introduce the term ``$\epsilon$-coherence length", where $\epsilon$ is the minimum required interference visibility (mutual coherence) and can in principle take on values between zero and one. In Fig. \ref{fig:res1} thus plot the cutoff for the $1/e$-coherence length, $1/2$-coherence length, and $7/8$ coherence length.  The coherence lengths obtained are summarized in table  
\begin{table}[]
    \centering
    \begin{tabular}{c|c|c|c|c}
        $I_{seed}$ ($\mu$W) &$\beta$ ($\%$) & $\epsilon={1}/{e}$ ($\mu$m) & $\epsilon=1/2$  ($\mu$m) & $\epsilon=7/8$  ($\mu$m)  \\
        \hline
         0 & 0 & 29.30 & 24.40 & 10.71 \\
          90 & $4.9 \pm 3.5$ & 30.64 & 25.32 & 11.00 \\
           270 & $13.9 \pm 2.7$& 33.74 & 27.32 & 11.61 \\
            540 & $20.5 \pm 2.5$& 36.88 & 29.17 & 12.12 \\
             3000 & $44.9 \pm 1.8$& $\infty$  & 45.30 & 14.87 \\
              10700 & $76.0 \pm 1.7$ & $\infty$  & $\infty$  & 25.12 \\
    \end{tabular}
    \caption{Normalized intensity of the stimulated component and $\epsilon$-coherence lengths obtained for different seed beam intensities.}
    \label{tab:1}
\end{table}
\par
We might look to other parameters in Eq. \eqref{l_t_stim} to characterize the  coherence length of a StimPDC light source. The  width of the gaussian contribution $\delta_{sp}$ is due to the spontaneous component alone, being the same for all values of $\beta$, and thus not a good parameter (except perhaps when $\beta$ is small, where the usual definition works fine).  The parameter $\beta$ gives the minimum coherence, and does not reveal information about the distance $d$ at which it can be reached.



 \section{Conclusion}
We have studied the coherence properties of the idler beam in StimPDC, with a view to measure the coherence length full idler beam, taking into account spontaneous and stimulated emission of light.  We showed theoretically that the idler beam is a weighted combination of two GSM beams, and thus is not a GSM beam in the general case.  This leads to a normalized mutual coherence function, that is not a simple gaussian distribution, which can render the usual definition of the coherence length meaningless in some cases.  We note that other experiments have shown non-trival and even non-monotonic coherence functions \cite{walborn11b,almeida12}.    Experimentally, we considered the case where the pump and seed laser beams have infinite coherence length. This configuration allows measuring and characterizing the effective coherence length of the idler as a function of the coherence length arising from the SPDC component and the normalized intensity of stimulation. We evaluated the visibility pattern as a function of the seed beam intensity, obtained in  a Michelson interferometer in which one path is displaced horizontally by a controlled length.   Motivated by quantum information applications, we introduce the notion of ``$\epsilon$-coherence length", defined as the largest distance between two interfering points of the field for which the visibility is above $\epsilon$, and use it to characterize the idler field.  Our work contributes to the development of new approaches for synthesizing partially coherent optical beams and introduces the notion of coherence length in non-trivial field distributions, according to its practical application. 

 \begin{acknowledgements}
This work was funded by the the Brazilian agencies Conselho Nacional de Desenvolvimento Cient\'{\i}fico e Tecnol\'ogico (CNPq - DOI 501100003593), Instituto Nacional de Ci\^encia e Tecnologia de Informa\c c\~ao Qu\^antica (INCT-IQ 465469/2014-0), Coordena\c c\~{a}o de Aperfei\c coamento de Pessoal de N\'\i vel Superior (CAPES DOI 501100002322), Funda\c c\~{a}o de Amparo \`{a} Pesquisa do Estado de Santa Catarina (FAPESC - DOI 501100005667), Funda\c c\~{a}o de Amparo \`{a} Pesquisa do Estado de Goi\'{a}s (FAPEG - DOI 501100005285),
and the Chilean agencies Fondo Nacional de Desarrollo Cient\'{i}fico y Tecnol\'{o}gico (FONDECYT - DOI 501100002850, regular project 1200266); National Agency of Research and Development (ANID), and  Millennium Science Initiative Program—ICN17-012.
\end{acknowledgements}

 \bibliographystyle{unsrt}
 \bibliography{StimPDC_CL}


\end{document}